\documentclass[aps,prc,twocolumn,showpacs,floatfix,longbibliography]{revtex4-1}

\usepackage[dvipdfmx]{graphicx}
\usepackage{hyperref} 
\usepackage{amssymb,amsmath,color} 
\usepackage{bm}

\begin{document}

\preprint{APS/123-QED}

\title{
  Colossal enhancement of spin-chirality-related Hall effect by thermal fluctuation
}

\author{Yasuyuki Kato}
\affiliation{
Department of Applied Physics, The University of Tokyo, Bunkyo, Tokyo, 113-8656, JAPAN 
}

\author{Hiroaki Ishizuka}
\affiliation{
Department of Applied Physics, The University of Tokyo, Bunkyo, Tokyo, 113-8656, JAPAN 
}

\date{\today}

\begin{abstract}
The effect of thermal fluctuation on the spin-chirality-induced anomalous Hall effect in itinerant magnets is theoretically studied. 
Considering a triangular-lattice model as an example, we find that a multiple-spin scattering induced by the fluctuating spins increases the Hall conductivity at a finite temperature. 
The temperature dependence of anomalous Hall conductivity is evaluated by a combination of an unbiased Monte Carlo simulation and a perturbation theory. 
Our results show that the Hall conductivity can increase up to $10^3$ times the ground state value; we discuss that this is a consequence of a skew scattering contribution. 
This enhancement shows the thermal fluctuation significantly affects the spin-chirality-related Hall effect. 
Our results are potentially relevant to the thermal enhancement of anomalous Hall effect often seen in experiments.
\end{abstract}


\maketitle

\section{Introduction}
Anomalous Hall effect (AHE) has been one of the central topics in the study of quantum transport phenomena~\cite{Hall1881}. Continuous study over more than a century have revealed that AHE shows rich properties which attracts the interest not only from basic science but also from applications (e.g., high-accuracy Hall-effect sensors)~\cite{Nagaosa2010,Maekawa2006}. Microscopically, the mechanism of the AHE is often classified into two groups: Intrinsic mechanism related to the Berry curvature of the electronic bands~\cite{Karplus1954} and the extrinsic mechanism due to impurity scattering~\cite{Smit1955,Smit1958,Berger1970}. The difference in the microscopic origin is often reflected in the behaviors of the AHE. For instance, the intrinsic AHE reflects singular structures in the Berry curvature. This gives rise to non-monotonic temperature ($T$)~\cite{Fang2003} and field~\cite{Takahashi2018} dependence of the anomalous Hall conductivity $\sigma_{\rm AHE}$. On the other hand, the extrinsic AHE by magnetic scatterings shows 
a peaklike enhancement of the Hall resistivity at a certain $T$ which characterizes the underlying physics such as the magnetic transition~\cite{Kondo1962} and coherence~\cite{Fert1987,Yamada1993,Kontani1994} $T$s. These rich features of the AHE have been intensively studied in both theory and experiment, and are also useful in identifying the physics behind the phenomena.

\begin{figure}[!htb]
  \centering
  \includegraphics[width= \columnwidth, trim =50 115 220 18, clip]{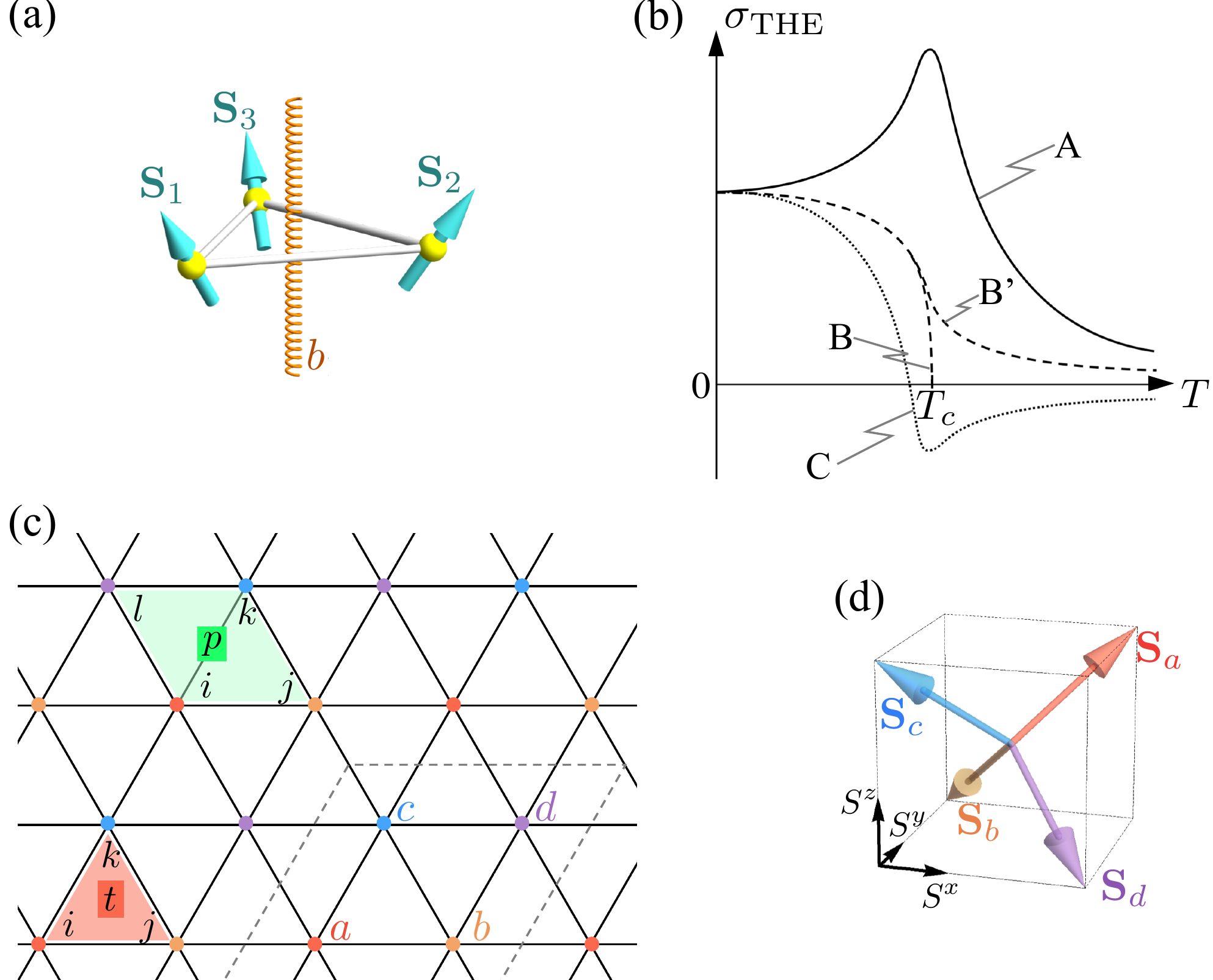}
  \caption{
  	Noncoplanar magnetic structure and topological Hall effect. (a) A schematic figure of the effective magnetic field
	 induced by a noncoplanar magnetic structure. 
	(b) Schematic figure for the $T$ dependence of anomalous Hall conductivity $\sigma_{\rm THE}$. 	
	Curve B' (B) indicates the case where $\sigma_{\rm THE}$ is proportional to the (spontaneous) scalar spin chirality
	while C indicates the case with extrinsic topological Hall effect in chiral magnets~\cite{Ishizuka2018}.
	In contrast, $\sigma_{\rm THE}$ in this study shows a different feature (Curve A). 
	(c) Triangular lattice model. The orange triangle (green plaquette) depicts the three-(four-) spins consisting $\chi_p$ ($h_p$).  
	Red, orange, blue, and purple sites indicate $a$-, $b$-, $c$-, and $d$-sublattice structure of the $3Q$ order, respectively. 
	(d) Spin orientation at each sublattice. All the spins in $\nu$-sublattice align to ${\bf S}_\nu$ at $T=0$. 
  }
  \label{fig1}
\end{figure}

Among various studies, a recent breakthrough was the discovery of AHE related to scalar spin chirality which is often called topological Hall effect (THE). Scalar spin chirality is a quantity defined by the scalar triple product of magnetic moments ${\bf S}_1 \cdot ({\bf S}_2 \times {\bf S}_3)$, where ${\bf S}_i$ is a local magnetic moment [Fig.~\ref{fig1}(a)]). This quantity is a measure of the non-coplanar nature of spin texture because the spin chirality is zero whenever the three spins lie in a same plane. It was pointed out that the spins produce a fictitious magnetic field $b$ when the three adjacent spins have a finite scalar spin chirality, resulting in an AHE~\cite{Ye1999,Ohgushi2000,Taguchi2001} [Fig.~\ref{fig1}(a)]. Alternatively, it is interpreted as an AHE due to the magnetic scattering by multiple scatterers~\cite{Tatara2002,Ishizuka2018}. The spin-chirality-related mechanism is studied in various materials, such as perovskite~\cite{Matl1998,Jakob1998,Chun2000,LyandaGeller2001} and pyrochlore~\cite{Taguchi2001} oxides, chiral magnets~\cite{Neubauer2009,Kanazawa2011,Yokouchi2014,Franz2014}, triangular oxides~\cite{Martin2008,Akagi2010,Kato2010,Takatsu2010,Ok2013}, and kagome antiferromagnets~\cite{Chen2014,Nakatsuji2015}. The THE in these materials are often investigated by the magnetic field dependence, which are consistent with the theoretical predictions~\cite{Taguchi2001,Neubauer2009,Kanazawa2011}.

In contrast, the $T$ dependence of the THE in the non-coplanar magnetic states is less understood. 
The Hall conductivity is expected to decrease with increasing temperature in magnets with non-coplanar magnetic orders because the scalar spin chirality decreases  [Curve B in Fig.~\ref{fig1}(b)]. In experiment, however, many materials show an increase of the Hall conductivity with increasing $T$~\cite{Takatsu2010,Ok2013,Yokouchi2014}
[Curve A of Fig.~\ref{fig1}(b)]; some materials show the maximum slightly above the magnetic transition temperature $T_c$~\cite{Takatsu2010,Ok2013}. This is in contrast to the known theories, where the maximum is expected to be below~\cite{Kondo1962} or much~\cite{Fert1987} higher than $T_c$. To the best of our knowledge, no theoretical understanding on the $T$ dependence is reached so far.

In this work, we theoretically study the enhancement of Hall conductivity ($\sigma_{\rm THE}$) by the thermal fluctuation focusing on the fluctuation-induced skew scattering. As an example, we consider a triangular lattice model with four-sublattice non-coplanar order called $3Q$ order [Figs.~\ref{fig1}(c) and \ref{fig1}(d)]. 
The $T$ dependence of $\sigma_{\rm THE}$ is calculated combining a Monte Carlo (MC) simulation and a large-size numerical calculation using Kubo formula. We find that $\sigma_{\rm THE}$ increases with increasing $T$, sometimes up to $10^3$ times compared with the ground state. The scan over the carrier density $n_{\rm el}$ $(0.1\leq n_{\rm el} \leq1.9)$, which is the average number of electrons per site, shows the enhancement due to the skew scattering by multiple spins generally appears in this model. 
Our results show the thermal fluctuation causes enhancement of AHE at finite $T$.

\section{Model}
In this study, we consider a classical Heisenberg spin model on a triangular lattice as an example of short-range non-coplanar magnetic order~\cite{Momoi1997}. The Hamiltonian reads
\begin{eqnarray}
\mathcal{H}_{\rm spin}&=&
K\sum_p h_p + D_4 \sum_i \left[ (S^x_i)^4 +  (S^y_i)^4 +  (S^z_i)^4 \right] \nonumber\\
&&-B_a\sum_{i\in a} \frac{S^x_i+S^y_i+S^z_i}{\sqrt{3}}
- B_{\chi} \sum_t \chi_t ,
\label{eq:model}
\end{eqnarray}
where 
${\bf S}_i$ ($|{\bf S}_i| = 1$) represents localized spin at site $i$, 
the sums $\sum_p$, $\sum_i$, and $\sum_t$ 
run over all the four-site plaquettes, all the sites, and all the three-site plaquettes, respectively, and 
$K$, $D_4$, $B_a$, and $B_\chi$ represent a short range multispin interaction, a single-ion anisotropy, a sublattice specific magnetic field, and a fictitious field coupled to the spin chirality, respectively. For each triangular plaquette $t$ and rhombic plaquette $p$,
multispin interactions are defined as
\begin{align}
&h_{p=\{i,j,k,l\}} \equiv ({\bf S}_i\cdot {\bf S}_j )({\bf S}_k\cdot {\bf S}_l )
-({\bf S}_i\cdot {\bf S}_l )({\bf S}_j\cdot {\bf S}_k ) \nonumber\\
&-({\bf S}_i\cdot {\bf S}_k )({\bf S}_j\cdot {\bf S}_l )\nonumber\\
&+({\bf S}_i\cdot {\bf S}_j + {\bf S}_i\cdot {\bf S}_k + {\bf S}_i\cdot {\bf S}_l 
+{\bf S}_j\cdot {\bf S}_k + {\bf S}_j\cdot {\bf S}_l + {\bf S}_k\cdot {\bf S}_l), \nonumber \\
&\chi_{t=\{i,j,k\}} \equiv {\bf S}_i \cdot ({\bf S}_j \times {\bf S}_k ),\nonumber
\end{align}
where
$\{i,j,k\}$ of a triangular plaquette $t$ is defined in order of counter-clockwise,
and $\{i,j,k,l\}$ of a rhombic plaquette $p$ is defined so that the pairs of $(i,k)$ and $(j,l)$
are the diagonal pairs of the corners of a rhombic plaquette [see Fig.~\ref{fig1}(c)].
A model with first $K$ term was originally introduced in the study on two-dimensional solid $^3$He~\cite{Thouless1965}. More recently, the biquadratic terms in $h_p$ was discussed in the effective spin models for the Kondo lattice model~\cite{Akagi2012,Ishizuka2015}.
The model with only $K$ terms ($D_4=B_a=B_\chi = 0$) exhibits a finite $T$ phase transition with the spontaneous $Z_2$ symmetry breaking from paramagnets to a chiral phase where spins are disordered but $\chi$s are ordered~\cite{Momoi1997}.

With $D_4$, $B_a$, and $B_\chi$, the low $T$ phase becomes a magnetic order because these terms reduce the symmetry: 
the $D_4$ term represents the single spin cubic anisotropy because of which spins favor one of $[\pm 1, \pm 1, \pm 1]$ directions ($D_4>0$);
the $B_a$ term represents the Zeeman coupling between the $a$-sublattice spins and an external magnetic field ${\bf B}_a\parallel [111]$ ($B_a>0$),
and the $B_\chi$ term represents a coupling between a fictitious field $B_\chi > 0$ and $\chi_t$ because of which $\chi > 0$ in entire $T$ range. 
With these three terms, the low-$T$ chiral phase is replaced by a four-sublattice long-range magnetic ordered phase [Figs.~\ref{fig1}(c) and \ref{fig1}(d)].
In this state, the four spins on each sublattice in Fig. 1(c) points along different directions [Fig. 1(d)], forming a non-coplanar magnetic texture.

\section{Results}
\noindent
{\it Monte Carlo simulation} --- 
\begin{figure}[!htb]
  \centering
  \includegraphics[width=\columnwidth,clip]{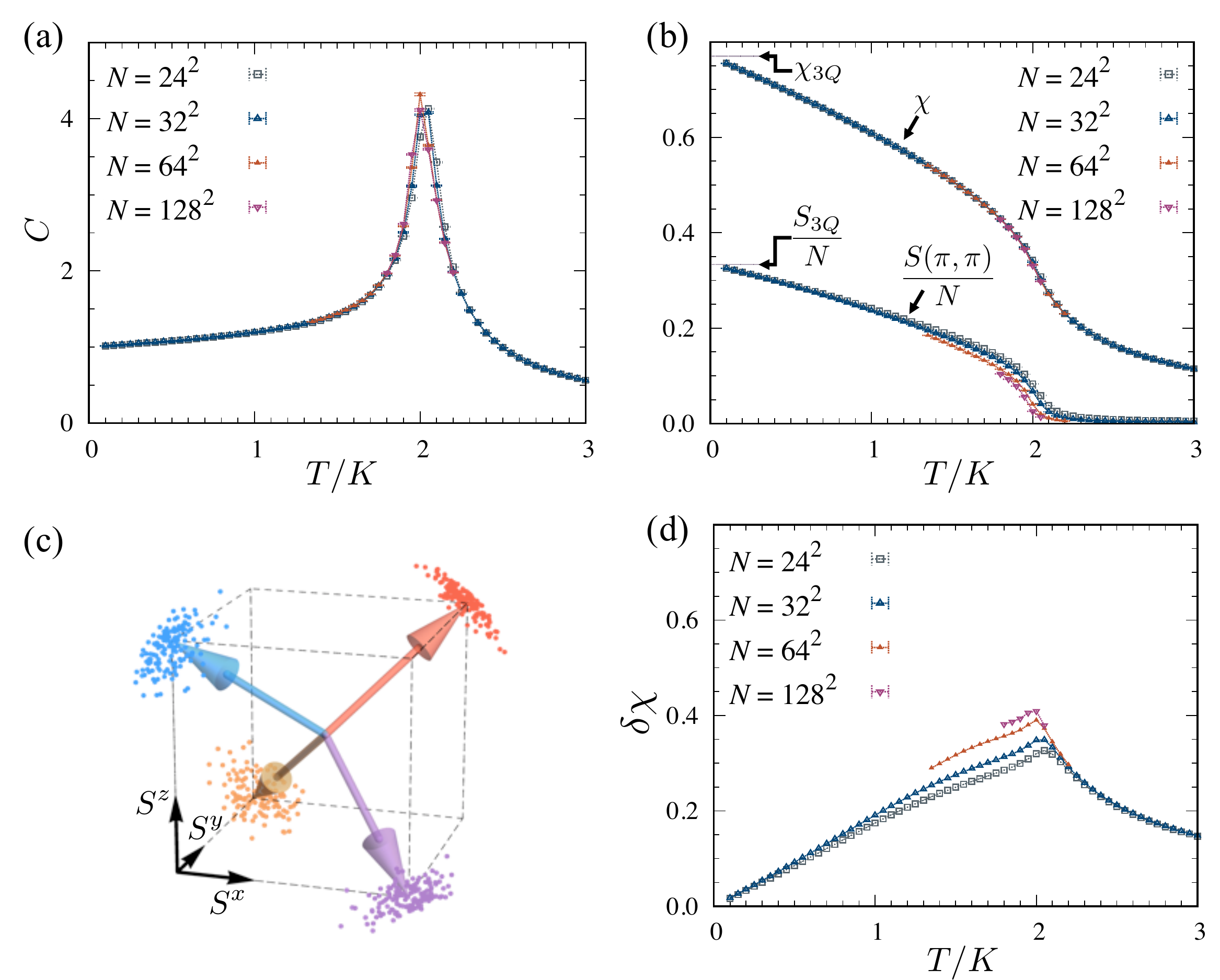}
  \caption{
  	Results of MC simulations of the model \eqref{eq:model}:
  	(a) specific heat $C$ and (b) normalized spin structure factor $S(\pi,\pi)/N$ and  scalar spin chirality $\chi$.
  	(c) A spin configuration in MC simulation at $T/K=0.1$.
  	Each point represents spin orientation.  Arrows are guide for eyes.
  	(d) Spin chirality of fluctuating spins $\delta \chi$. 
       }
  \label{fig2}
\end{figure}
The finite $T$ properties of this model is calculated by MC simulations using a standard single-spin-flip Metropolis algorithm~\cite{supp}. 
Figure~\ref{fig2} shows the results of MC simulations with $\mathcal{H}_{\rm spin}$. The specific heat $C$ in Fig.~\ref{fig2}(a) shows a peak at $T_c/K \simeq 2$, and the normalized structure factor 
$S(\pi,\pi)/N
=\langle [ ( \sum_{i \in a,c} {\bf S}_i )- ( \sum_{i \in b,d} {\bf S}_i ) ]^2  \rangle /N^2$, [$N(=L^2)$ is the number of spins and $L$ is the size of the system.]
becomes nonzero below $T_c$ reflecting a phase transition to a magnetic order phase. 
In the lowest $T$,
$S(\pi,\pi)/N$ approaches $S_{3Q}/N = 1/3$,  and
$\chi$ approaches
$\chi_{3Q} = 4/(3\sqrt{3})$ as shown in Fig.~\ref{fig2}(b).
Figure~\ref{fig2}(c) shows a spin configuration obtained in simulation at the sufficiently low $T$. 
These results consistently show that the ground state is the $3Q$ order and the phase transition is continuous. 
We also find that the overall behavior of the above quantities are sufficiently converged with $L \geq 24$ with some finite size effect close to $T_c$. 
The behavior of $C$, $S(\pi,\pi)/N$, and $\chi$ as well as the observed finite size effect indicates that the phase transition is continuous.

We note that the scalar spin chirality remains positive in the entire $T$ range, even above $T_c$ [Fig.~\ref{fig2}(b)]. 
The nonzero $\chi$ comes from the local correlation of fluctuating spins
under the $B_\chi$ field in Eq.~\eqref{eq:model}, which acts as the ``magnetic field'' for $\chi$. 
As a measure of the chirality due to the fluctuating spins, we use
\begin{align}
\delta\chi=\frac{\chi}{\chi_{3Q}} - \left[ \frac{S(\pi,\pi)}{S_{3Q}} \right]^{\frac{3}{2}}.\label{eq:Xfluc}
\end{align}
In contrast,  $\delta\chi$ shows a different $T$ dependence. 
Figure~\ref{fig2}(d) shows $\delta\chi$ increases with increasing $T$ and shows a cusp like peak at $T_c$. 
The magnetic scattering by fluctuating spins produces anomalous Hall effect proportional to the scalar spin chirality~\cite{Tatara2002,Ishizuka2018}. Therefore, the fluctuating spins may produce a non-monotonic $T$ dependence of $\sigma_{\rm THE}$.

\noindent
{\it Anomalous Hall conductivity} --- To study the $T$ dependence of $\sigma_{\rm THE}$, 
we consider itinerant electrons coupled to the spins in ${\cal H}_{\rm spin}$. 
The electrons are coupled to the localized spins via Hund's coupling, 
i.e., we consider a Kondo lattice model on the triangular lattice. The Hamiltonian reads:
\begin{equation}
\mathcal{H}_{\rm KL}  = 
-t \hspace{-1mm} \sum_{\langle i,j\rangle, s}
(c^\dag_{is} c^{\;}_{js} + {\rm h.c.})
 -J  \hspace{-1mm}\sum_{i,s,s'} {\bf S}_{i} \cdot ( c^\dag_{is} {\bm \sigma}_{ss'} c^{\;}_{is'}  ),
\label{eq:KLM}
\end{equation}
where ${\bm \sigma}=(\sigma^x, \sigma^y, \sigma^z)$ are the vector of Pauli matrices, and
$c^\dag_{is}$ ($c^{\;}_{is}$) is a creation (annihilation) operator of itinerant electron at site $i$ with spin $s$.
The first term represents the kinetic energy term of itinerant electrons, and the second term the Hund's coupling.
We assume that the coupling is relatively weak ($J=t/2$), and the energy scale in the electron system is much larger than the spin system. 
Then, for simplicity, we fix the temperature of the electron system $T_{\rm el}/ t = 0.025$.
The Hall conductivity $\sigma_{\rm THE}$ is calculated by Kubo formula using spin configurations generated by the MC simulation~\cite{
Yi2009,
Ueland2012,
Ishizuka2013,
Ishizuka2013b,
Rosales2018,
supp}.
\begin{figure}[!htb]
  \centering
 \includegraphics[trim=95 75 165 20, clip,width=\columnwidth]{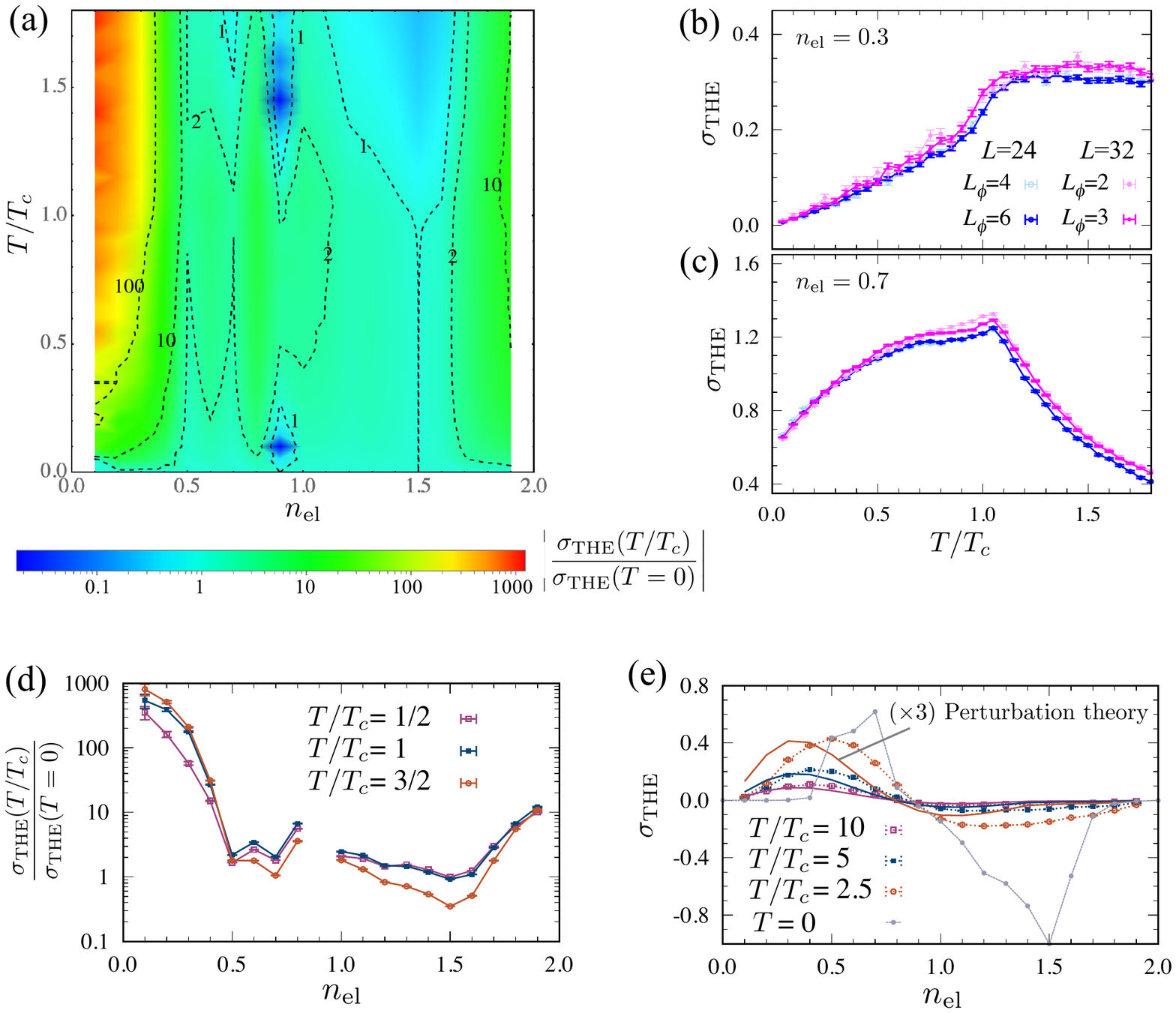}
  \caption{
  Hall conductivity $\sigma_{\rm THE}$ as a function of $n_{\rm el}$ and $T$. 
  (a) Contour plot of $| \sigma_{\rm THE}(T/T_c)/\sigma_{\rm THE}(T=0)|$ computed with $L=32$ and $L_\phi = 3$. 
  (b,c) $\sigma_{\rm THE}(T/T_c)$ at $n_{\rm el} = $ 0.3 and 0.7.   
  (d) Ratio of $\sigma_{\rm THE}(T/T_c=\text{1/2, 1, or 3/2})$ and $\sigma_{{\rm THE}}(T=0)$~\cite{fn1}.
  (e) $n_{\rm el}$ dependence of 
  $\sigma_{\rm THE}( T \gg T_c)$ and $\sigma_{\rm THE} (T = 0)$. 
  Solid lines are obtained by the perturbation theory with a scale factor 3 for visibility~\cite{Tatara2002}. 
  }
  \label{fig3}
\end{figure}

Figures~\ref{fig3}(b,c) show $\sigma_{\rm THE}(T)$ at  $n_{\rm el} =\text{0.3 and 0.7}$ as examples. 
Different lines are for different choices of $L$ and $L_\phi$; 
we find only small finite size effect after taking the average over the twisted boundaries. 
When $n_{\rm el}=0.3$ [Fig.~\ref{fig3}(b)], we find $\sigma_{\rm THE}\sim 10^{-2}e^2/h$ at $T/T_c=0.05\ll 1$. 
However, $\sigma_{\rm THE}$ monotonically increases with increasing $T$, reaching $\sigma_{\rm THE}\sim0.3e^2/h$ at $T\sim T_c$; 
this is approximately 30 times larger than $\sigma_{\rm THE}(T/T_c =0.05)$. 
In our calculation, we find the enhancement of $\sigma_{\rm THE}$ in nearly all choices of $n_{\rm el}$. 
Figure~\ref{fig3}(c) shows the results for $n_{\rm el}=0.7$. 
Similar to Fig.~\ref{fig3}(b), $\sigma_{\rm THE}$ increases with increasing $T$ and decreases above $T_c$; 
the curve shows a maximum around $T\sim T_c$ with a kink slightly below it. 
This trend also appears for $n_{\rm el}>1$ except the difference in the sign of $\sigma_{\rm THE}$~\cite{supp}.

The increase of $\sigma_{\rm THE}$ implies the enhancement is related to the fluctuation effect. 
Indeed, the $T$ dependence of $\sigma_{\rm THE}$ is in contrast to that of $\chi$, which decreases monotonically with increasing $T$ [Fig.~\ref{fig2}(b)]. 
Therefore, the enhancement is different from what is expected in the intrinsic THE mechanism. 
On the other hand, the increase of $\sigma_{\rm THE}$ below $T_c$ and the maximum around $T_c$ are coincident with the $T$ dependence of $\delta\chi$.
Furthermore, at some filling e.g., $n_{\rm el}=0.7$ [Fig.~\ref{fig3}(c)],  $\sigma_{\rm THE}$ shows a cusp at $T_c$ resembling $\delta\chi$.
These features imply the enhancement is related to the spin chirality of fluctuating spins $\delta\chi$, presumably related to the skew scattering mechanism~\cite{Ishizuka2018} 

Our results in Fig.~\ref{fig3}(a) also find that the thermal effect is larger when the Fermi level is close to the band edge, i.e., $n_{\rm el}\sim 0$ or $\sim 2$. 
Figure~\ref{fig3}(d) shows the ratio $\sigma_{\rm THE}(T)/\sigma_{\rm THE}(T=0)$ at $T/T_c=$1/2, 1, and 3/2.
As shown in the figure, 
$\sigma_{\rm THE}(T/T_c=1/2, 1, 3/2)$ is typically 2--10 times larger than $\sigma_{\rm THE}(T=0)$.
On the other hand, the enhancement at the band edges are much larger, sometimes up to $10^3$ times of that at $T=0$. 

The $n_{\rm el}\sim 0$ and $\sim 2$ regions are close to the ideal setup in which the skew scattering is often studied. In the case $n_{\rm el}\sim 0$ ($n_{\rm el}\sim 2$), the electron (hole) bands are well approximated by the quadratic dispersion. The skew scattering in these situations are often driven by the large-angle scattering, which is dominant when $\xi k_F\ll 1$.
Here, $k_F$ and $\xi$ are the Fermi wavenumber and the correlation length for $\chi_t$, respectively. 
Therefore, the skew scattering theory in Ref.~\cite{Ishizuka2018} applies to this case.
The $\xi k_F\ll 1$ condition is not satisfied when the chemical potential moves away from the band edge.
Hence, the results may generally change due to the large Fermi surface. Nevertheless, our result shows the enhancement is commonly seen regardless of the size of the Fermi surface.

\noindent
{\it High-temperature region} --- We next turn to the $n_{\rm el}$ dependence of $\sigma_{\rm THE}$ at a high-$T$ region well above $T_c$. 
The results of MC simulation are shown in Fig.~\ref{fig3}(e). The results show a qualitatively different $n_{\rm el}$ dependence compared to the $T=0$ result; 
$\sigma_{\rm THE}$ for $n_{\rm el}<1$ tends to be larger than that for $n_{\rm el}>1$ in the high $T$ regime while the trend is opposite in the low $T$ near $T=0$.
This contrasting trend at a high-$T$ is explained by the relaxation-time (electron lifetime) dependence of skew scattering mechanism.
To see the density of state [$\rho(\mu)$] dependence, we evaluated $\sigma_{\rm THE}$ using a perturbation method in Ref.~\cite{Tatara2002}. In the $T\gg T_c$ region, the fluctuation contribution is expected to be the only contribution to the Hall effect. Also, the correlation length of the spins becomes very short in this region. Therefore, we only take into account the contribution from the nearest-neighbor spin correlation. With these approximations, the conductivity reads~\cite{Tatara2002}:
\begin{align}
\sigma_{\rm THE}
=&-\frac{e^2J^3\tau^2}{\pi N}\sum_{(ijk)\in t}\epsilon_{\alpha\beta\gamma}\langle S^\alpha_{{\bf r}_k}S^\beta_{{\bf r}_i}S^\gamma_{{\bf r}_j}\rangle \nonumber\\
&\qquad\times I_x({\bf r}_j - {\bf r}_k )I_0({\bf r}_k - {\bf r}_i )I_y({\bf r}_i - {\bf r}_j),\label{eq:sigma_pert}
\end{align}
where
$
I_a({\bf r})\equiv \frac1{\tau N}\sum_{\bf k}\frac{v_{\bf k}^a e^{{\rm i}{\bf k}\cdot {\bf r} }}{\varepsilon_{{\bf k}}^2+1/(4\tau^2)},
$
and ${\bf v}_{\bf k}$ ($\varepsilon_{\bf k}$) is the velocity (energy) of electrons with momentum ${\bf k}$ ($v^0_{\bf k} = 1$). 
Here, the sum of $(ijk)$ is limited to the three spins forming the triangles $t$ [Fig.~\ref{fig1}(c)]. 
The electron lifetime $\tau$ is evaluated using the first Born approximation, $\tau^{-1}(\mu)=2\pi J^2\rho(\mu)$. 
Here, we neglected the spin-spin correlation for the evaluation of $\tau$. 

The result of Eq.~\eqref{eq:sigma_pert} is shown in Fig.~\ref{fig3}(e). 
The perturbation theory semi-quantitatively reproduces the overall trend of numerical results.
The similarity between the numerical results and the perturbation suggests that the Hall effect is related to the skew scattering by the fluctuating spins in the high-$T$ regime;
in the perturbation theory, larger skew scattering contribution to $\sigma_{\rm THE}$ is expected when 
 $\tau \propto \rho(\mu)$ is larger~\cite{Nagaosa2010,Tatara2002,Ishizuka2018}, and indeed $\rho(\mu)$ for $n_{\rm el} <1$ is smaller than that for $n_{\rm el}>1$.

\noindent
\section{Summary and concluding remarks} 
To summarize, in this work, we studied the effect of the thermal fluctuation to the spin-chirality-related anomalous Hall effect. 
By an unbiased numerical simulation, we find the Hall conductivity $\sigma_{\rm THE}$ increases with increasing temperature, sometimes approximately $10^3$ times the ground state value.
Detailed analysis on the temperature and electron-density dependence shows the enhancement is consistent with the skew scattering mechanism proposed recently~\cite{Ishizuka2018}; 
the thermal enhancement is larger when the Fermi level is close to the band edge, and is also related to the density of states. 
These results show a significant effect of the thermal fluctuation to the Hall effect induced by non-coplanar magnetic orders.

In contrast to our results, the skew scattering mechanism was also discussed in relation to the sign change of $\sigma_{\rm THE}$ close to the critical temperature in chiral magnets with long-period magnetic orders (e.g., MnGe)~\cite{Ishizuka2018}. 
This is a decidedly different behavior from the current case where the skew scattering enhances the Hall effect. 
Presumably, a key difference is the size of the magnetic structure, i.e., the characteristic wave number $k^*$ is large (small) in the $3Q$ order (magnetic skyrmion crystals). 
In the skew scattering mechanism~\cite{Ishizuka2018}, the scattering amplitude is proportional to $\sin\theta$ where $\theta$ is the angle between the in-comming and out-going electrons, namely, larger angle scattering is important. 
In addition to the skew scattering, the small angle scattering is also induced by the intrinsic topological Hall effect (THE) when $k^*$ is small.
In other words, from the scattering theory viewpoint, the scattering channels for the skew scattering and intrinsic THE are different for small $k^*$.
In contrast, since the magnetic unit cell of $3Q$ order has only four sites ($k^*$ is large), both the skew scattering and intrinsic THE induce a large angle scattering. 
Our results presented here shows that the magnetic fluctuation plays a non-trivial and crucial role in magnets with such a short period order.

\begin{acknowledgments}
The authors thank J. M. Ok, Y. Motome, and N. Nagaosa for fruitful discussions. This work was supported by JSPS KAKENHI Grant Numbers JP16H02206, JP16H06717, JP18K03447, JP18H03676, JP18H04222, JP19K14649, and JP26103006, 
and CREST, JST (Grant Nos. JPMJCR16F1, JPMJCR18T2, and JPMJCR1874).
Numerical calculations were conducted on the supercomputer system in ISSP, The University of Tokyo.
\end{acknowledgments}

\bibliography{tahe.bib}

\begin{thebibliography}{46}
\bibitem [{fn1()}]{fn1}%
  \BibitemOpen
  \href@noop {} {}\bibinfo {note} {The results for $n_{\rm el}=0.9$ are not
  displayed because of large numerical errors.}\BibitemShut {Stop}%
\end{thebibliography}
\end{document}